\documentclass[preprint,aps,amsmath,showpacs,showkeys,nofootinbib,
superscriptaddress]{revtex4}
\usepackage[dvips]{graphicx}

\begin{document}

\title{\bf Multi-$\bar K$ hypernuclei}

\author{D.~Gazda}
\email{gazda@ujf.cas.cz}
\affiliation{Nuclear Physics Institute, 25068 \v{R}e\v{z}, Czech Republic}

\author{E.~Friedman}
\email{elifried@vms.huji.ac.il}
\affiliation{Racah Institute of Physics, The Hebrew University, 
Jerusalem 91904, Israel}

\author{A.~Gal}
\email{avragal@vms.huji.ac.il}
\affiliation{Racah Institute of Physics, The Hebrew University, 
Jerusalem 91904, Israel}

\author{J.~Mare\v{s}}
\email{mares@ujf.cas.cz}
\affiliation{Nuclear Physics Institute, 25068 \v{R}e\v{z}, Czech Republic}

\date{\today}

\begin{abstract}

Relativistic mean-field calculations of multi-$\bar K$ \emph{hypernuclei} 
are performed by adding $K^-$ mesons to particle-stable configurations of 
nucleons, $\Lambda$ and $\Xi$ hyperons. For a given hypernuclear core, 
the calculated $\bar K$ separation energy $B_{\bar K}$ saturates with the 
number of $\bar K$ mesons for more than roughly 10 mesons, with $B_{\bar K}$ 
bounded from above by 200~MeV. The associated baryonic densities saturate at 
values $2-3$ times nuclear-matter density within a small region where the 
$\bar K$-meson densities peak, similarly to what was found for multi-$\bar K$ 
nuclei. The calculations demonstrate that particle-stable multistrange 
$\{ N,\Lambda,\Xi \}$ configurations are stable against strong-interaction 
conversions $\Lambda \to N {\bar K}$ and $\Xi \to N {\bar K} {\bar K}$, 
confirming and strengthening the conclusion that kaon condensation is 
unlikely to occur in strong-interaction self-bound strange hadronic matter. 

\end{abstract} 



\maketitle 

\newpage

\section{Introduction} 
\label{sec:intro} 

Quasibound nuclear states of $\bar K$ mesons have been studied by us recently in
a series of articles \cite{MFG05,MFG06,GFGM07,GFGM08}, using a self-consistent
extension of nuclear relativistic mean-field (RMF) models.
References \cite{MFG05,MFG06,GFGM07} focused on the widths expected for $\bar
K$ quasibound states, particularly in the range of $\bar K$ separation energy
$B_{\bar K} \sim 100-150$ MeV deemed relevant from $K^-$-atom phenomenology
\cite{MFG06,FGa07} and from the KEK-PS E548 $^{12}{\rm C}(K^-,N)$ missing-mass
spectra \cite{kish07} that suggest values of Re~$V_{\bar K}(\rho_0) \sim
-(150-200)$ MeV. Such deep potentials are not reproduced at present by chirally
based approaches that yield values of Re~$V_{\bar K}(\rho_0)$ of order $-100$
MeV or less attractive, as summarized recently in Ref.~\cite{WHa08}. For a
recent overview of $\bar K N$ and $\bar K$-nucleus dynamics, see
Ref.~\cite{galexa08}. 

The subject of multi-$\bar K$ nuclei was studied in 
Refs.~\cite{GFGM07,GFGM08}, where the focal question considered was 
whether or not kaon condensation could occur in strong-interaction 
self-bound nuclear matter. Yamazaki {\it et al.}~\cite{YDA04} argued 
that $\bar{K}$ mesons might provide the relevant physical degrees of 
freedom for reaching high-density self-bound strange matter that
could then be realized as multi-$\bar{K}$ nuclear matter. This scenario 
requires that $B_{\bar{K}}$ beyond some threshold value of strangeness 
exceeds $m_Kc^2 + \mu_N - m_{\Lambda}c^2 \gtrsim 320$~MeV, where $\mu_N$ 
is the nucleon chemical potential, thus allowing for the conversion 
$\Lambda \to \bar{K} + N$ in matter. For this strong $\bar K$ binding, 
$\Lambda$ and $\Xi$ hyperons would no longer combine with nucleons to compose 
the more conventional kaon-free form of strange hadronic matter, which is made 
out of $\{ N,\Lambda,\Xi \}$ particle-stable configurations \cite{SDG93,SDG94} 
(see Ref.~\cite{SBG00} for an update), and $\bar K$ mesons would condense then 
macroscopically. However, our detailed calculations in Ref.~\cite{GFGM08} 
demonstrated a robust pattern of saturation for $B_{\bar{K}}$ and for nuclear 
densities upon increasing the number of $\bar K$ mesons embedded in the 
nuclear medium. For a wide range of phenomenologically allowed values 
of meson-field coupling constants compatible with assuming a deep 
$\bar K$-nucleus potential, the saturation values of $B_{\bar K}$ were found 
generally to be below 200 MeV, considerably short of the threshold value of
$\approx 320$ MeV required for the onset of kaon condensation under laboratory 
conditions. Similar results were subsequently published by Muto {\it et al.} 
\cite{MMT09}. Our discussion here concerns kaon condensation in self-bound 
systems, constrained by the strong interactions. It differs from discussions 
of kaon condensation in neutron stars where weak-interaction constraints are 
operative for any given value of density. For very recent works on kaon 
condensation in neutron-star matter, see Ref.~\cite{BGB08}, where hyperon 
degrees of freedom were disregarded, and Ref.~\cite{Muto08a}, where the 
interplay between kaon condensation and hyperons was studied, and references 
to earlier relevant work cited therein. 

In our calculations of multi-$\bar{K}$ nuclei \cite{GFGM08}, the saturation 
of $B_{\bar{K}}$ emerged for any boson-field composition that included the 
dominant vector $\omega$-meson field, using the F-type SU(3) value $g_{\omega 
KK}\approx 3$ associated with the leading-order Tomozawa-Weinberg term of the 
meson-baryon effective Lagrangian. This value is smaller than in any of the 
other commonly used models \cite{GFGM08}. Moreover, the contribution of each 
one of the vector $\phi$-meson and $\rho$-meson fields was found to be 
substantially repulsive for systems with a large number of antikaons, 
reducing $B_{\bar{K}}$ as well as lowering the threshold value of the number 
of antikaons required for saturation to occur. We also verified that the 
saturation behavior of $B_{\bar K}$ is qualitatively independent of the RMF 
model applied to the nucleonic sector. The onset of saturation was found to 
depend on the atomic number. Generally, the heavier the nucleus is, the more 
antikaons it takes to saturate their separation energies. We concluded that 
$\bar{K}$ mesons do not provide a constituent degree of freedom for self-bound 
strange dense matter. 

In the present work we extend our previous RMF calculations of multi-$\bar{K}$ 
nuclei into the domain of multi-$\bar{K}$ \emph{hypernuclei}, to 
check whether a joint consideration of $\bar{K}$ mesons together with hyperons 
could bring new features or change our previous conclusions. This is the first 
RMF calculation that considers both $\bar{K}$ mesons and hyperons together 
within {\it finite self-bound} hadronic configurations. The effect of 
hyperonic strangeness in bulk on the dispersion of kaons and antikaons 
was considered by Schaffner and Mishustin \cite{SMi96}. More recently, 
kaon-condensed hypernuclei as highly dense self-bound objects have been 
studied by Muto \cite{Muto08b}, using liquid-drop estimates. 

The plan of the article is as follows. In Sec.~\ref{sec:model} we briefly 
outline the RMF methodology for multi-$\bar{K}$ hypernuclei and discuss 
the hyperon and ${\bar K}$ couplings to the meson fields used in the present 
work. Results of these RMF calculations for multi-$\bar{K}$ hypernuclei are 
shown and discussed in Sec.~\ref{sec:res}. We conclude with a brief summary 
and outlook in Sec.~\ref{sec:fin}.

\section{Model} 
\label{sec:model} 

\subsection{RMF formalism} 

In the present work, our interest is primarily aimed at multiply strange
baryonic systems containing (anti)kaons. We employed the relativistic mean-field
approach where the strong interactions among pointlike hadrons are mediated by
\emph{effective} mesonic degrees of freedom. In the following calculations we
started from the Lagrangian density
\begin{equation} 
\label{eq:l} 
\begin{split} 
{\mathcal L} &=\bar{B}\left[{\rm i}\gamma^\mu D_\mu 
-(M_B-g_{\sigma B}\sigma-g_{\sigma^* B}\sigma^*) \right]B \\ 
&+\left( D_\mu K \right)^\dagger 
\left( D^{\,\mu} K \right)-(m_K^2-g_{\sigma K}\,m_K\sigma-
g_{\sigma^* K}\,m_K\sigma^*)K^\dagger K \\ 
&+ (\sigma,\sigma^*,\omega_\mu,\vec{\rho}_\mu,\phi_\mu,A_\mu \, 
\text{free-field terms}) -U(\sigma)-V(\omega), 
\end{split}
\end{equation} 
which includes, in addition to the common isoscalar scalar ($\sigma$), isoscalar
vector ($\omega$), isovector vector ($\rho$), electromagnetic ($A$) fields, and
nonlinear self-couplings $U(\sigma)$ and $V(\omega)$, also \emph{hidden
strangeness} isoscalar $\sigma^*$ and $\phi$ fields that couple exclusively to
strangeness degrees of freedom. Vector fields are coupled to baryons $B$
(nucleons, hyperons) and $K$ mesons via the covariant derivative
\begin{equation} 
D_\mu = \partial_\mu 
+ {\rm i}\, g_{\omega \Phi}\, \omega_\mu + {\rm i}\, g_{\rho \Phi}\, 
\vec{I} \cdot \vec{\rho}_\mu + {\rm i}\, g_{\phi \Phi}\, \phi_\mu 
+ {\rm i}\, e\, (I_3+ \textstyle\frac{1}{2}\displaystyle Y) A_\mu \:, 
\end{equation} 
where $\Phi=B$ and $K$, with $\vec{I}$ denoting the isospin operator, 
$I_3$ being its $z$ component, and $Y$ standing for hypercharge. 
This particular choice of the coupling scheme for $K^-$ mesons ensures the 
existence of a conserved Noether current, the timelike component of which 
can then be normalized to the number of $K^-$ mesons in the medium,
\begin{equation} 
\rho_{K^-}=2 
(E_{K^-}+g_{\omega K}\,\omega+g_{\rho K}\,\rho+g_{\phi K}\,\phi+e\, A) 
K^+K^-, \qquad \int {\rm d}^3 x\, \rho_{K^-} = \kappa , 
\end{equation} 
and serves as a dynamical source in the equations of motion for the boson 
fields in matter: 
\begin{equation} 
\begin{split} 
(-\nabla^2 + m_\sigma^2)\sigma =& 
\:g_{\sigma B} \bar{B}B + g_{\sigma K} m_K K^+\hspace{-1pt} K^- 
-\frac{\partial}{\partial \sigma} U(\sigma)  \\ 
(-\nabla^2 \hspace{-2pt} + m_\sigma^{*2})\sigma^*\hspace{-4pt} =& 
\:g_{\sigma^* B} \bar{B}B + g_{\sigma^* K} m_K K^+\hspace{-1pt} K^-  \\ 
(-\nabla^2 + m_\omega^2) \omega =& 
\:g_{\omega B} B^\dagger \hspace{-1pt}B - g_{\omega K} \rho_{K^-} 
+\frac{\partial}{\partial \omega} V(\omega) \\ 
(\,-\nabla^2 + m_\rho^2) \rho =& 
\:g_{\rho B} B^\dagger I_3 B - g_{\rho K} \rho_{K^-}  \\ 
(-\nabla^2 + m_\phi^2)\phi =& \:g_{\phi B} B^\dagger \hspace{-1pt}B 
- g_{\phi K} \rho_{K^-}  \\ 
-\nabla^2 A =& \:e\, B^\dagger(I_3+\textstyle\frac{1}{2}\displaystyle Y)B 
- e\, \rho_{K^-}. 
\end{split} 
\end{equation} 
These \emph{dynamically} generated intermediate fields then enter the Dirac 
equation for baryons,
\begin{equation} 
\left[ -{\rm i} 
\mbox{\boldmath$\alpha$\unboldmath}\cdot\mbox{\boldmath$\nabla$\unboldmath} 
+\beta \left( M_B-g_{\sigma B}\sigma-g_{\sigma^* B}\sigma^* \right) 
+g_{\omega B}\omega 
+g_{\rho B}I_3\rho 
+g_{\phi B}\phi 
+e\left(I_3+\textstyle\frac{1}{2}\displaystyle Y \right)A 
\right]B=\epsilon B 
\end{equation} 
and the Klein-Gordon equation for $K^-$ mesons,
\begin{equation} 
\label{eq:Kkg} 
[-\mbox{\boldmath$\nabla$\unboldmath}^2-E_{K^-}^2 +m_K^2 + \Pi_{K^-} ]K^-=0, 
\end{equation} 
with the in-medium $K^-$ self-energy,
\begin{equation} 
\begin{split} 
\Pi_{K^-}=&-g_{\sigma K}m_K\sigma-g_{\sigma^* K}m_K\sigma^* 
-2E_{K^-}(g_{\omega K}\omega+g_{\rho K}\rho+g_{\phi K}\phi+eA) \\ 
&-(g_{\omega K}\omega+g_{\rho K}\rho+g_{\phi K}\phi+eA)^2. 
\end{split} 
\end{equation} 
Hence, the presence of the $\bar{K}$ mesons modifies the scalar and vector mean
fields entering the Dirac equation, consequently leading to a \emph{dynamical}
rearrangement of the baryon configurations and densities that, in turn, modify
the ${\bar K}$ quasibound  states in the medium. This requires a self-consistent
solution of these coupled wave equations, a procedure followed numerically in
the present as well as in our previous works. In the present work, for the sake
of simplicity, we have suppressed the imaginary part of $\Pi_{K^-}$ arising from
in-medium $K^-$ absorption processes except for demonstrating its effect in one
example. Note that, for the range of values $B_{K^-} \gtrsim 100$ MeV mostly
considered here, the effect of Im~$\Pi_{K^-}$ was found to be negligible (see
Fig.~1 of Ref.~\cite{GFGM08}).

\subsection{Choice of the model parameters} 

To parametrize the nucleonic part of the Lagrangian density (\ref{eq:l}) 
we considered the standard RMF parameter sets NL-SH \cite{SNR93} and NL-TM1(2) 
\cite{STo94}, which have been successfully used in numerous calculations of 
various nuclear systems. 

In the case of hyperons the coupling constants to the vector fields were 
fixed using SU(6) symmetry. For $\Lambda$ hyperons this leads to 
\begin{equation} 
\label{eq:lambda} 
g_{\omega \Lambda}=\frac{2}{3}g_{\omega N}, \,\, g_{\rho \Lambda}=0, \,\, 
g_{\phi \Lambda}=\frac{-\sqrt{2}}{3}g_{\omega N}. 
\end{equation} 
The coupling to the scalar $\sigma$ field, $g_{\sigma \Lambda}/g_{\sigma 
N}=0.6184\, (0.623)$ for the NL-SH (NL-TM) RMF model, was then estimated by 
fitting to measured $\Lambda$-hypernuclear binding energies~\cite{HTa06}. This 
essentially ensures the well depth of 28~MeV for $\Lambda$ in nuclear matter. 
The coupling of the $\Lambda$ hyperon to the scalar $\sigma^*$ field was fixed 
by fitting to the measured value $\Delta B_{\Lambda\Lambda}\approx1$~MeV of 
the uniquely identified hypernucleus $_{\Lambda\Lambda}^{~~6}{\rm He}$ 
\cite{Tak01}. For $\Xi$ hyperons, SU(6) symmetry gives
\begin{equation} 
\label{eq:xi} 
g_{\omega\Xi}=\frac{1}{3}g_{\omega N}, \,\, g_{\rho \Xi}=-g_{\rho N}, \,\, 
g_{\phi\Xi}=-2\frac{\sqrt{2}}{3}g_{\omega N}. 
\end{equation} 
Because there are no experimental data for $\Xi(\Lambda)$-$\Xi$ interactions, 
we set $g_{\phi\Xi}=g_{\sigma^* \Xi}=0$ to avoid parameters that might lead to 
unphysical consequences and that, in addition, are expected to play a minor 
role (in analogy to the small effect, of order 1 MeV for $B_{K^-}$, found upon 
putting $g_{\phi\Lambda}$ and $g_{\sigma^*\Lambda}$ to zero, and as is 
demonstrated below in Fig.\ 7 within a different context). The coupling to 
the scalar $\sigma$ field was then constrained to yield an optical potential 
Re~$V_{\Xi^-}=-14$~MeV in the center of $^{12}$C \cite{Kha00}. This 
corresponds to $g_{\sigma \Xi}=0.299 g_{\sigma N}$ for the NL-TM2 RMF model. 

\begin{table} 
\caption{$\bar K$ and $K^-$ separation energies, $B_{\bar K}$ and 
$B_{K^-}$, respectively, calculated statically (in MeV) for a single antikaon 
$1s$ state in several nuclei, using the NL-TM nuclear RMF parametrizations 
(TM2 for $^{12}$C and $^{16}$O, TM1 for $^{40}$Ca and above) and vector SU(3) 
coupling constants, Eq.~(\ref{eq:KSU(3)}). The difference $B_{K^-}-B_{\bar K}$ 
is due to the $K^-$ finite-size Coulomb potential.} 
\label{tab:t1} 
\begin{ruledtabular} 
\begin{tabular}{lccccc} 
 & $^{12}$C & $^{16}$O & $^{40}$Ca & $^{90}$Zr & $^{208}$Pb \\ 
\hline 
$B_{\bar K}$ & 44.8 & 42.7 & 49.8 & 54.5 & 53.6 \\ 
$B_{K^-}$ & 49.0& 47.6 & 59.2 & 69.4 & 76.6 \\ 
\end{tabular} 
\end{ruledtabular} 
\end{table} 

Finally, for the antikaon couplings to the vector meson fields we adopted 
a purely F-type, vector SU(3) symmetry: 
\begin{equation} 
\label{eq:KSU(3)} 
2g_{\omega K}=2g_{\rho K} = \sqrt{2}g_{\phi K}=g_{\rho \pi}=6.04, 
\end{equation} 
where $g_{\rho\pi}$ is due to the $\rho\rightarrow 2\pi$ decay width 
\cite{WHa08}. (Here we denoted by $g_{VP}$ the VPP electric coupling 
constant $g_{VPP}$.) Using this ``minimal'' set of coupling constants 
to establish correspondence with chirally based approaches, 
we calculate the single antikaon $1s$ separation energies $B_{\bar K}$ and 
$B_{K^-}$ listed in Table~\ref{tab:t1}. These separation energies are lower 
roughly by 25~MeV than those anticipated from $\bar K N - \Sigma \pi$ 
coupled-channel chiral approaches \cite{WHa08}, most likely because the 
$K^{\star}$ vector-meson off-diagonal coupling is not included in the standard 
RMF formulation. The missing attraction, and beyond it, is incorporated 
here by coupling the antikaon to scalar fields $\sigma$ and $\sigma^*$. 
SU(3) symmetry is not of much help when fixing the coupling constants of 
scalar fields. Because there still is no consensus about the microscopic origin 
of the scalar $\sigma$ field and the strength of its coupling to ${\bar K}$ 
mesons \cite{TKO08,KMN09}, in this work we fitted $g_{\sigma K}$ to several 
assumed $K^-$ separation energies $B_{K^-}$ in the range of $100-150$ MeV 
for a single $K^-$ meson in selected nuclei across the periodic table, 
as implied by the deep $K^-$-nucleus potential phenomenology of 
Refs.~\cite{MFG06,kish07}. Furthermore, for use in multistrange 
configurations, the coupling constant to the $\sigma^*$ field is taken 
from $f_0(980)\rightarrow K{\bar K}$ decay to be $g_{\sigma^* K}=2.65$ 
\cite{SMi96}. The effect of the $\sigma^*$ field was found generally to be 
minor. For a more comprehensive discussion of the issue of scalar couplings, 
see our previous work \cite{GFGM08}.

\subsection{Inclusion of the SU(3) baryon octet} 

We considered many-body systems consisting of the SU(3) octet 
$N,\Lambda,\Sigma$, and $\Xi$ baryons that can be made particle stable 
against strong interactions \cite{SDG93,SDG94}. The energy release $Q$ values 
for various conversion reactions of the type $B_1B_2\rightarrow B_3B_4$ 
together with phenomenological guidance on hyperon-nucleus interactions 
suggest that only the conversions $\Xi^-p\rightarrow \Lambda\Lambda$ and 
$\Xi^0n\rightarrow \Lambda\Lambda$ (for which $Q\simeq 20$~MeV) can be 
overcome by binding effects. It becomes possible then to form particle-stable 
multi-$\{ N,\Lambda,\Xi \}$ configurations for which the conversion 
$\Xi N\rightarrow\Lambda\Lambda$ is Pauli blocked owing to the $\Lambda$ 
orbitals being filled up to the Fermi level. For composite configurations with 
$\Sigma$ hyperons the energy release in the $\Sigma N \rightarrow \Lambda N$ 
conversion is too high ($Q\gtrsim 75$~MeV) and, hence, it is unlikely for 
hypernuclear systems with $\Sigma$ hyperons to be particle stable.

\section{Results and discussion} 
\label{sec:res} 

In Refs.~\cite{GFGM07,GFGM08} we studied multi-$\bar K$ nuclei, observing 
that the calculated $K^-$ separation energies as well as the nuclear densities 
saturate upon increasing the number of $K^-$ mesons embedded dynamically in 
the nuclear medium. This saturation phenomenon, which is qualitatively 
independent of the applied RMF model, emerged for any boson-field composition 
containing the dominant vector $\omega$-meson field which acts repulsively 
between $\bar K$ mesons. Because the calculated $K^-$ separation energies did 
not exceed 200~MeV, for coupling-constant combinations designed to bind 
a single $K^-$ meson in the range $B_{K^-} \sim 100-150$ MeV, it was argued 
that kaon condensation is unlikely to occur in strong-interaction self-bound 
hadronic matter. In this section we demonstrate that these conclusions hold 
also when adding, within particle-stable multistrange configurations, large 
numbers of hyperons to nuclei across the periodic table.

\subsection{Multi-$\{ N, \Lambda, K^- \}$ configurations} 

\begin{figure}
\includegraphics[scale=0.7]{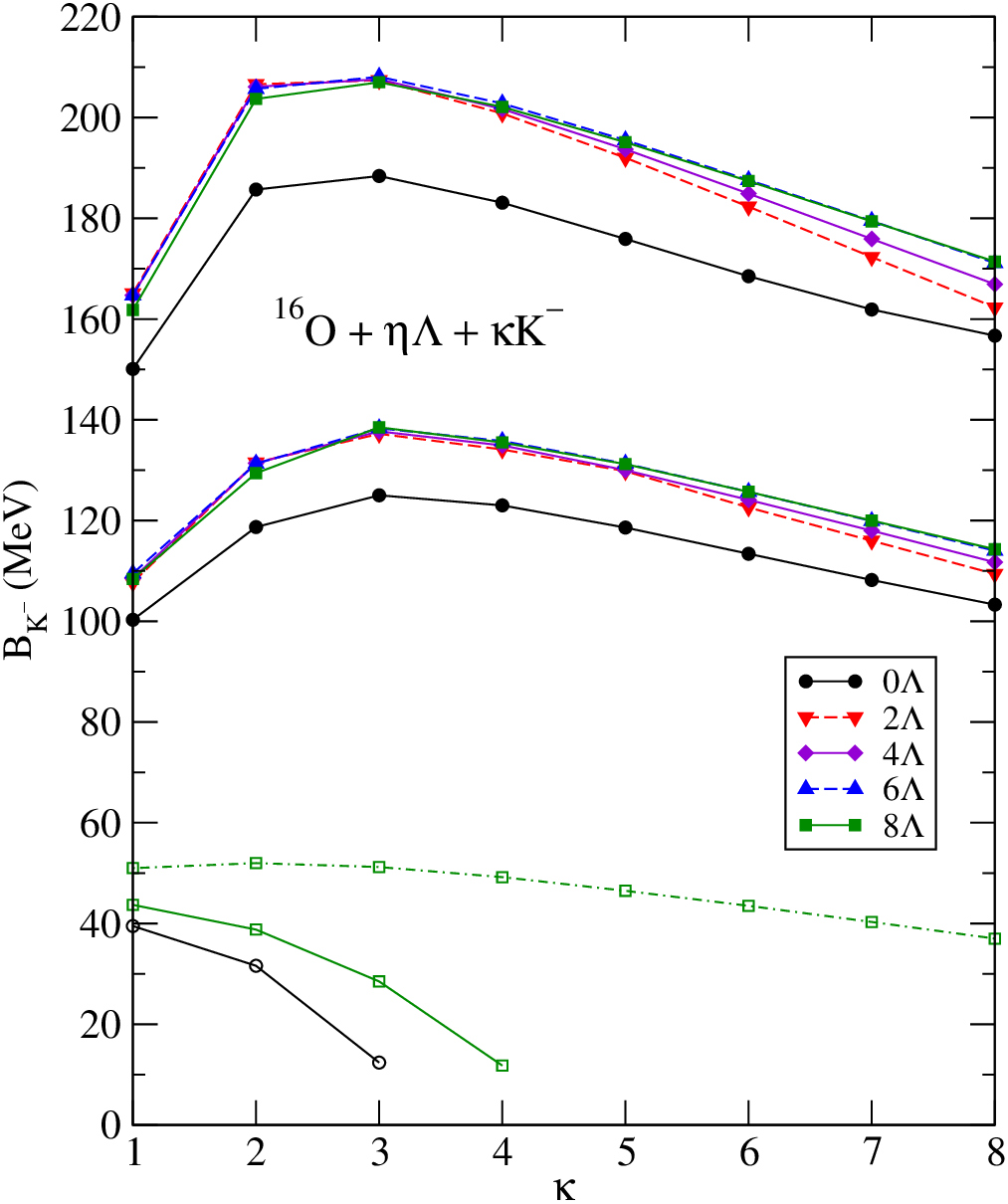} 
\caption{(Color online) The $1s$ $K^-$ separation energy $B_{K^-}$ in 
$^{16}{\rm O}+\eta\Lambda +\kappa K^-$ as a function of the number $\kappa$ of 
antikaons for several values of the number $\eta$ of $\Lambda$ hyperons, with 
initial values $B_{K^-}=100$ and 150~MeV for $\eta=0$, $\kappa =1$, calculated 
in the NL-SH RMF model. The solid (dot-dashed) lines with open symbols 
correspond to $g_{\sigma K}=0$ including (excluding) Im~$\Pi_{K^-}$.} 
\label{fig:f1} 
\end{figure}

Figure~\ref{fig:f1} presents $1s$ $K^-$ separation energies 
$B_{K^-}$ in $^{16}{\rm O}+\eta \Lambda +\kappa K^-$ 
multi-$K^-\Lambda$ hypernuclei as a function of the number 
$\kappa$ of $K^-$ mesons for $\eta = 0,2,4, 6$, and 8 $\Lambda$ 
hyperons, calculated in the NL-SH model for two values of $g_{\sigma K}$ 
($g_{\sigma K}=0.233 g_{\sigma N}$ and $0.391 g_{\sigma N}$) chosen to 
produce $B_{K^-}=100$ and 150~MeV, respectively, for $\eta=0$, $\kappa=1$. 
In addition, the lower group of curves with $B_{K^-}<60$~MeV corresponds to 
$g_{\sigma K}=0$. The figure illustrates saturation of 
$B_{K^-}$ with the number of antikaons in multi-$\Lambda$ 
hypernuclei. There is an apparent increase of $B_{K^-}$ (up to 
15\%) when the first two $\Lambda$ hyperons fill the $1s$ shell. 
Further $\Lambda$ hyperons, placed in the $p$ shell, cause only 
insignificant variation of $B_{K^-}$ for small values of $\kappa$. 
However, the effect of the $1p_{3/2}$-shell hyperons increases 
with the number of antikaons, and for $\kappa=8$ it adds another 
$5-10$~MeV to $B_{K^-}$. The separation energy $B_{K^-}$ remains 
almost unaffected (or even decreases) by the next two $\Lambda$ 
hyperons placed in the $1p_{1/2}$ shell. The figure thus suggests 
saturation of the $K^-$ separation energy also with the number 
$\eta$ of $\Lambda$ hyperons in the nuclear medium. When the $K^-$ 
coupling to the $\sigma$ field is switched off, $g_{\sigma K}=0$, 
the $K^-$ separation energy assumes relatively low values, 
$B_{K^-}\lesssim 50$~MeV, and decreases as a function of $\kappa$ 
when Im~$\Pi_{K^-}$ is considered (solid lines). In this case, the 
effect of $K^-$ absorption is not negligible as illustrated by the 
dot-dashed line showing $B_{K^-}$ for Im~$\Pi_{K^-}=0$. 
The effect of Im~$\Pi_{K^-}\neq 0$ for $B_{K^-}>100$~MeV in the upper groups 
of curves is negligible and is not shown here or in all subsequent figures. 

It is worth noting that $\eta=8$ is the maximum number of $\Lambda$ 
hyperons in our calculation that can be bound in the $^{16}$O 
nuclear core. In some of the $^{16}{\rm O}+\eta\Lambda+\kappa K^-$ 
allowed configurtions, $1p_{1/2}$ neutrons became less bound than 
$1d_{5/2}$ neutrons because of the strong spin-orbit interaction. 
(This occurs, e.g., for $\eta=0$ when $\kappa\geq 5$ or for 
$\eta=8$ when $\kappa \geq 3$.) However, the total binding energy 
of the system was found always to be higher for configurations 
with $1p_{1/2}$ neutrons. Consequently, the standard shell 
configurations of oxygen are more bound and are thus energetically 
favorable. 

\begin{figure} 
\includegraphics[scale=0.6]{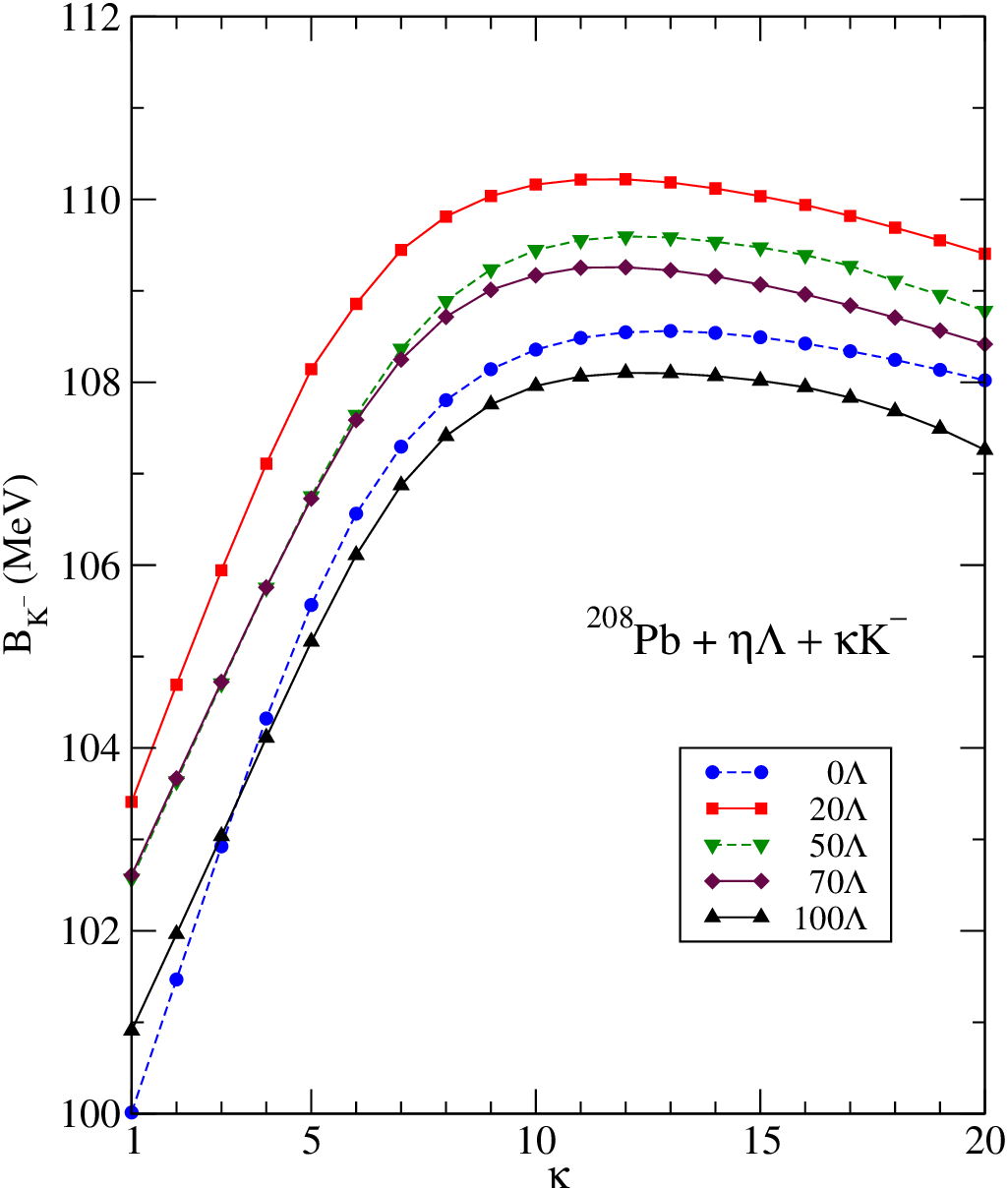} 
\caption{(Color online) The $1s$ $K^-$ separation energy $B_{K^-}$ in 
$^{208}{\rm Pb} + \eta \Lambda + \kappa K^-$ as a function of the number 
$\kappa$ of antikaons for several values of the number $\eta$ of $\Lambda$ 
hyperons, with initial value $B_{K^-}=100$~MeV for $\eta=0$, $\kappa = 1$, 
calculated in the NL-TM1 RMF model.} 
\label{fig:f2} 
\end{figure} 

The saturation of $B_{K^-}$ upon increasing the number of $\Lambda$ hyperons 
in multi-$K^-\Lambda$ hypernuclei based on a $^{16}$O nuclear core holds also 
when going over to heavier core nuclei. Figure~\ref{fig:f2} shows the $1s$ 
$K^-$ separation energy $B_{K^-}$ in $^{208}{\rm Pb}+\eta \Lambda +\kappa K^-$ 
multi-$K^-\Lambda$ hypernuclei as a function of both the number $\kappa$ of 
$K^-$ mesons and $\eta$ of $\Lambda$ hyperons, calculated in the NL-TM1 model 
for $g_{\sigma K}=0.133g_{\sigma N}$ such that $B_{K^-}=100$~MeV for $\eta=0$, 
$\kappa=1$. For any given number $\eta$ of $\Lambda$ hyperons, $B_{K^-}$ 
saturates with the number $\kappa$ of $K^-$ mesons, reaching its maximum value 
for $\kappa =12$. Morever, $B_{K^-}$ increases with the number of hyperons up 
to $\eta=20$, when it reaches its maximum value $B_{K^-} \approx 110$~MeV for 
$\kappa=12$, and then starts to decrease with $\eta$. Consequently, in the Pb 
configurations with 100 $\Lambda$ hyperons and more than 5 $K^-$ mesons, $K^-$ 
mesons are even less bound than in configurations with no $\Lambda$ hyperons. 
The decrease of $B_{K^-}$ with $\eta$ beyond $\eta=20$ is apparently related 
to a depletion of the central nuclear density in the presence of a massive 
number of hyperons in outer shells, as confirmed by some of the subsequent 
figures, because $B_{K^-}$ is greatly affected by the central nuclear density.

\subsection{Multi-$\{ N, \Lambda, \Xi, K^- \}$ configurations} 

When building up baryonic multi-$\{ N,\Lambda,\Xi \}$ 
configurations with maximum strangeness for selected core nuclei, 
we first started by filling up $\Lambda$ hyperon single-particle 
states in a given nuclear core up to the $\Lambda$ Fermi level. 
Subsequently, we added $\Xi^0$ and $\Xi^-$ hyperons as long as the reaction 
$[AN,\eta\Lambda,\mu\Xi]\rightarrow[(A-1)N,\eta\Lambda,(\mu-1)\Xi]+2\Lambda$ 
was energetically forbidden (here, $[...]$ denotes a bound configuration). 
Finally, we checked that the inverse reaction 
$[AN,\eta\Lambda,\mu\Xi]\rightarrow[(A+1)N,(\eta-2)\Lambda,(\mu+1)\Xi]$ 
is kinematically blocked as well. These conditions guarantee that 
such $\{ N,\Lambda, \Xi \}$ multistrange configurations are 
particle stable against strong interactions, decaying only via 
weak interactions. 

Clearly, the amount of $\Xi$ hyperons bound in a given system depends on 
the depth $-V_{\Xi}$ of the $\Xi$-nucleus potential. We adopted a value for 
$g_{\sigma\Xi}$ that gives $V_{\Xi}^{\rm Dirac}=V_{\rm S}+V_{\rm V}=-18$~MeV, 
corresponding to a depth of $-V_{\Xi}^{\rm Schr.}\simeq 14$~MeV for use in the 
Schroedinger equation \cite{Kha00}. For comparison, in some cases we also 
considered $V_{\Xi}^{\rm Dirac}=-25$~MeV. 

\begin{figure} 
\includegraphics[scale=0.7]{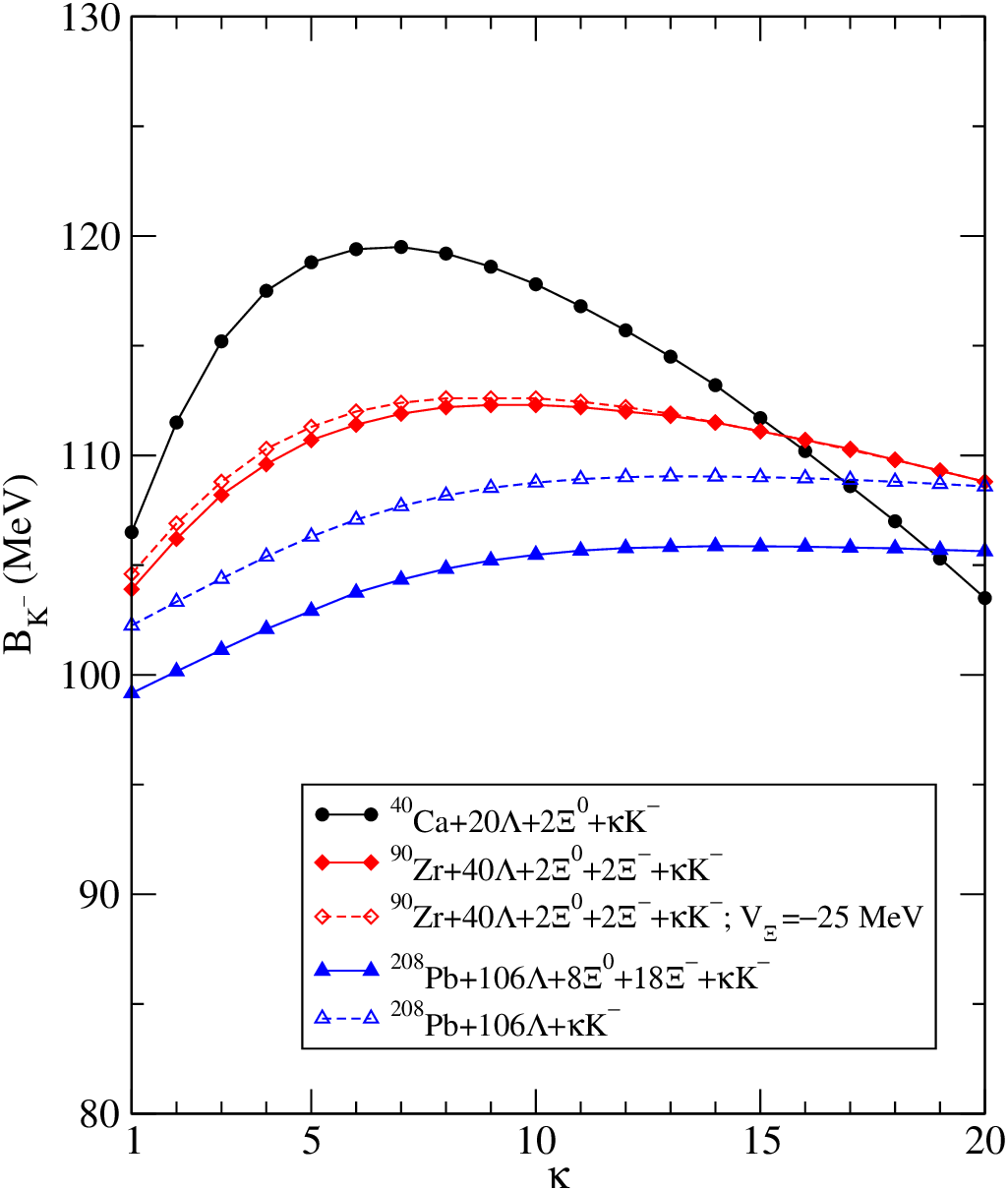} 
\caption{(Color online) The $1s$ $K^-$ separation energy $B_{K^-}$ in 
$^{40}{\rm Ca}$, $^{90}{\rm Zr}$, and $^{208}{\rm Pb}$ with $\eta\Lambda + 
\mu\Xi +\kappa K^-$ as a function of the number $\kappa$ of antikaons, 
with initial value $B_{K^-}=100$~MeV for $\eta=\mu=0$, $\kappa = 1$, 
calculated in the NL-TM1 RMF model.} 
\label{fig:f3} 
\end{figure} 

The $^{16}$O core can accommodate up to $\eta=8$ $\Lambda$ hyperons in 
particle-stable configurations, and the $^{16}{\rm O} + 8 \Lambda$ system 
admits many more, of order 40 $K^-$ mesons. However, we have not found any 
energetically favorable conversion $\Lambda\Lambda\rightarrow\Xi N$ in 
$^{16}{\rm O}+\eta\Lambda+\kappa K^-$ systems. Therefore, $\Xi$ hyperons 
are not part of any particle-stable multistrange configurations built upon 
the $^{16}$O core. While checking the energy balance in heavier systems 
with $^{40}$Ca, $^{90}$Zr, and $^{208}$Pb nuclear cores, 
we found particle-stable configurations: 
$^{40}{\rm Ca}+20\Lambda +2\Xi^0$, $^{90}{\rm Zr}+40\Lambda +2\Xi^0+2\Xi^-$, 
and $^{208}{\rm Pb}+106\Lambda +8\Xi^0+18\Xi^-$. We then embedded several 
$K^-$ mesons in these configurations and studied density distributions and 
binding energies in such multi-$K^-$ hypernuclear systems. Figure~\ref{fig:f3} 
demonstrates the calculated $1s$ $K^-$ separation energy $B_{K^-}$ in 
$^{40}{\rm Ca}+20\Lambda +2\Xi^0+\kappa K^-$, $^{90}{\rm Zr}+40\Lambda + 
2\Xi^0+2\Xi^-+\kappa K^-$, and $^{208}{\rm Pb}+106\Lambda +8\Xi^0+18\Xi^- 
+\kappa K^-$ as a function of the number $\kappa$ of $K^-$ mesons. 
For comparison, in the case of the $^{208}$Pb core, we also present calculations
done excluding $\Xi$ hyperons but keeping the same number, $\eta=106$, of
$\Lambda$ hyperons. A decrease of $B_{K^-}$ upon adding hyperons ($\Xi$ in this
case) is noted, in line with the trend observed and discussed for
Fig.~\ref{fig:f2} above. 

The calculations shown in Fig.~\ref{fig:f3} were performed within the NL-TM1 
nuclear RMF scheme using values of $g_{\sigma K}=0.211 g_{\sigma N}$ 
($^{40}$Ca) and $0.163 g_{\sigma N}$ ($^{90}$Zr), which yield $B_{K^-}=100$ MeV 
for a single $K^-$ nuclear configuration with $\eta=\mu=0$, where $\mu$ 
denotes the number of $\Xi$ hyperons. The figure demonstrates that the 
saturation of $K^-$ separation energies, observed for multi-$\Lambda$ 
hypernuclei in Figs.~\ref{fig:f1} and \ref{fig:f2}, holds also when $\Xi$
hyperons are added dynamically within particle-stable configurations and that
the heavier the system is, the larger number $\kappa$ of antikaons it takes to
saturate $B_{K^-}$. It is worth noting that in all cases $B_{K^-}$ does not
exceed 120~MeV. Finally, the two curves for a $^{90}$Zr nuclear core in
Fig.~\ref{fig:f3} (using diamond symbols) show the sensitivity to the value
assumed for the $\Xi$ hyperon potential depth, the standard $-V_{\Xi}^{\rm
Dirac}=18$ MeV, and a somewhat increased depth $-V_{\Xi}^{\rm Dirac}=25$~MeV,
illustrating the tiny effect it exercises on $B_{K^-}$ that is noticeable only
for $\kappa < 12$. 

\begin{figure} 
\includegraphics[scale=0.65]{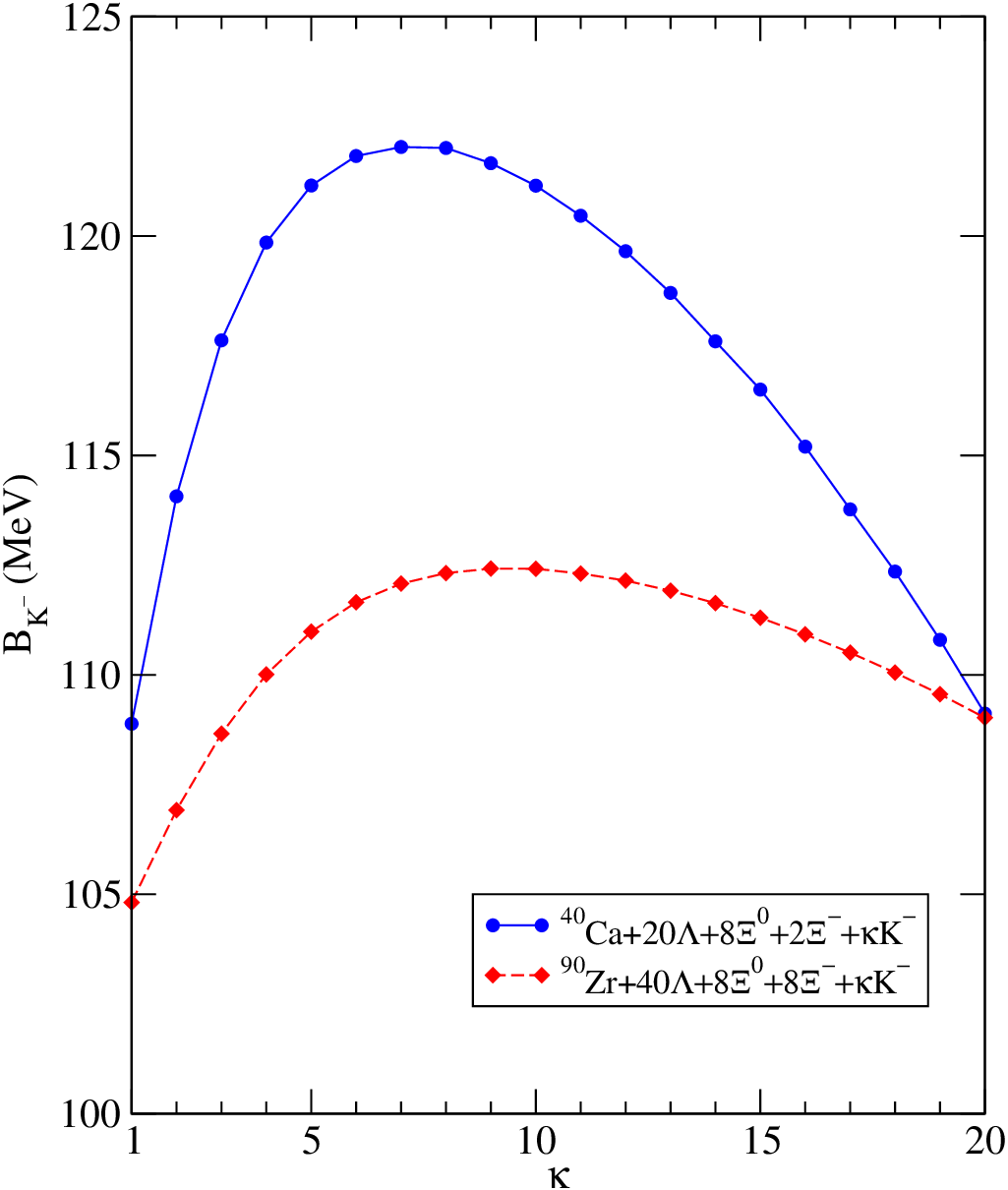} 
\caption{(Color online) The $1s$ $K^-$ separation energy $B_{K^-}$ in 
$^{40}{\rm Ca}$ and $^{90}{\rm Zr}$ with $\eta\Lambda +\mu\Xi +\kappa K^-$, 
for $V_{\Xi}^{\rm Dirac}=-25$~MeV, as a function of the number $\kappa$ of 
antikaons, with initial value $B_{K^-}=100$~MeV for $\eta=\mu=0$, $\kappa=1$, 
calculated in the NL-TM1 RMF model.} 
\label{fig:f4} 
\end{figure} 

A deeper $\Xi$ potential supports binding of more $\Xi$ hyperons in a given 
multi-$\Lambda$ hypernucleus. For $V_{\Xi}^{\rm Dirac}=-18$~MeV, only 2$\Xi^0$ 
and $2\Xi^0+2\Xi^-$ hyperons were found to be bound in $^{40}{\rm Ca}+ 
20\Lambda$ and $^{90}{\rm Zr}+40\Lambda$, respectively. However, 
for $V_{\Xi}^{\rm Dirac}=-25$~MeV it is possible to accommodate up to $8\Xi^0+
2\Xi^-$ hyperons in $^{40}{\rm Ca}+20\Lambda$ and $8\Xi^0+8\Xi^-$ hyperons in 
$^{90}{\rm Zr} + 40\Lambda$. Figure~\ref{fig:f4} presents the $1s$ $K^-$ 
separation energy $B_{K^-}$ in multi-$K^-$ hypernuclei 
$^{40}{\rm Ca} + 20\Lambda + 8\Xi^0 + 2\Xi^- + \kappa K^-$ and 
$^{90}{\rm Zr} + 40\Lambda + 8\Xi^0 + 8\Xi^- + \kappa K^-$ as a function 
of the number $\kappa$ of $K^-$ mesons, calculated in the NL-TM1 model for 
$V_{\Xi}^{\rm Dirac}=-25$ MeV, using values for $g_{\sigma K}$ such that 
$B_{K^-}=100$ MeV in $^{40}{\rm Ca} +1K^-$ and in $^{90}{\rm Zr} +1K^-$. 
The figure illustrates that the saturation of the $K^-$ separation energy 
occurs also in baryonic systems with three species of hyperons, $\Lambda$, 
$\Xi^0$, and $\Xi^-$, reaching quite large fractions of strangeness 
[${|S|}/{B} = 0.57(0.8)$ for a Ca(Zr) core]. We note that the separation 
energy $B_{K^-}$ barely exceeds 120~MeV in these cases too. 

\begin{figure} 
\includegraphics[scale=0.7]{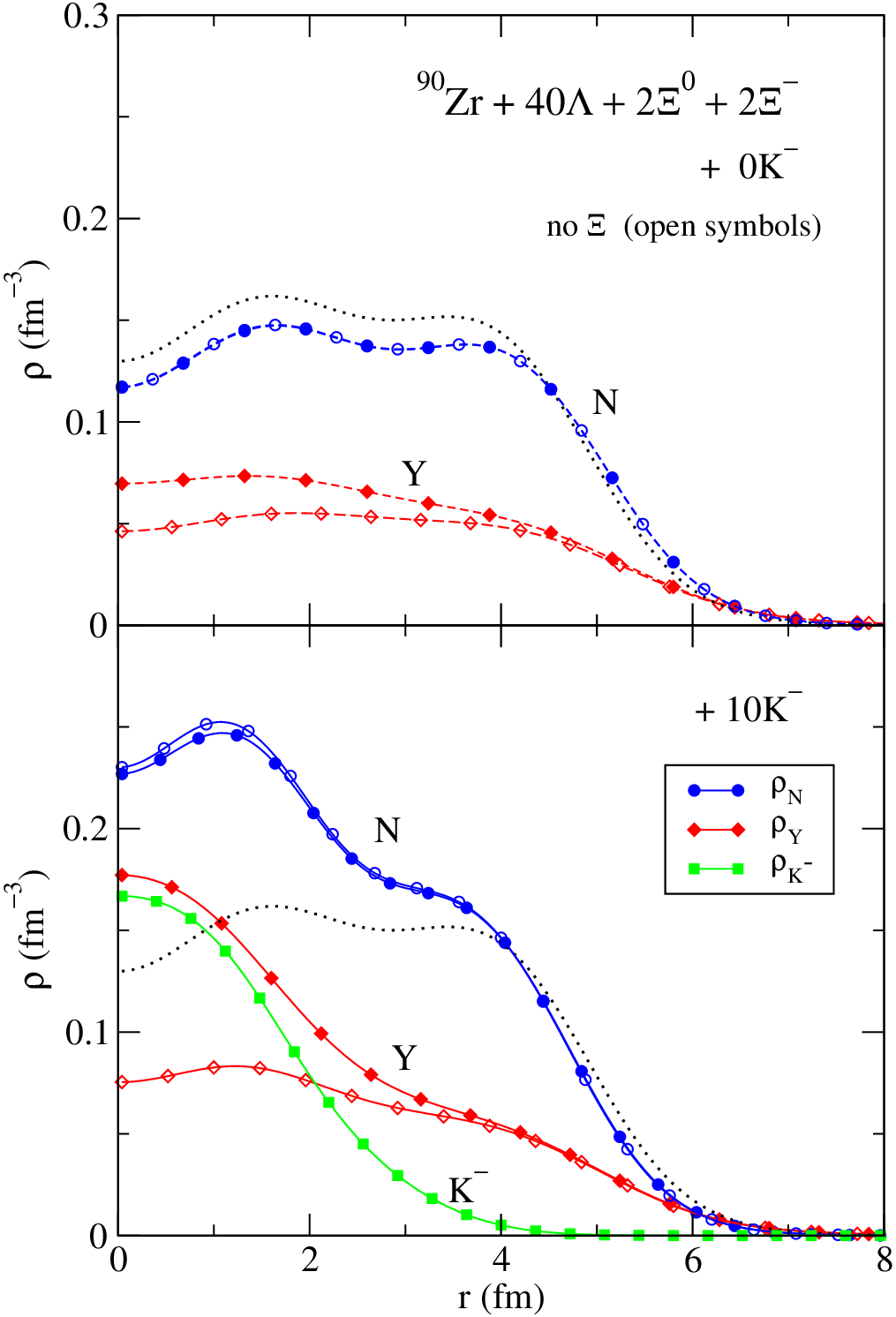} 
\caption{(Color online) Density distributions in $^{90}{\rm Zr}+40\Lambda 
+2\Xi^0 +2\Xi^- + \kappa K^-$, for $\kappa=0$ (top panel) and $\kappa=10$ 
(bottom panel), with $B_{K^-}=100$~MeV for $\eta=\mu=0$, $\kappa = 1$, 
calculated in the NL-TM1 RMF model. The dotted line corresponds to the nucleon 
density $\rho_N$ in $^{90}$Zr. The densities $\rho_{\Lambda}$ (open diamonds) 
and $\rho_N$ (open circles) in $^{90}{\rm Zr} + 40\Lambda + \kappa K^-$ are 
shown for comparison.} 
\label{fig:f5} 
\end{figure} 

We also studied the rearangement of nuclear systems induced by embedding 
hyperons and $K^-$ mesons. Figure~\ref{fig:f5} presents the evolution of the 
density distributions in Zr after first adding $40\Lambda+4\Xi$ hyperons (top 
panel) and then 10 $K^-$ mesons (bottom panel). The nucleon density $\rho_N$ 
in $^{90}$Zr is denoted by a dotted line. The relatively weakly bound hyperons 
with extended density distributions (dashed line, solid diamonds) attract 
nucleons, thus depleting the central nucleon density $\rho_N$ (dashed line, 
circles). Adding extra 10 $K^-$ mesons to the hypernuclear system induces 
large rearrangement of the baryons. The $K^-$ mesons, which pile up near the 
origin (solid line, squares), attract the surrounding nucleons and hyperons. 
Consequently, the densities $\rho_N$ and $\rho_Y$ (solid lines, solid circles 
and diamonds, respectively) increase considerably in the central region. The 
resulting configuration $^{90}{\rm Zr}+40\Lambda +2\Xi^0+2\Xi^- +10K^-$ is 
thus significantly compressed, with central baryon density $\rho_B$ exceeding 
the nuclear density in $^{90}$Zr by a factor of roughly 3. 

For comparison we present in Fig.~\ref{fig:f5} also the $\Lambda$ hyperon 
($\rho_{\Lambda}$, open diamonds) and nucleon ($\rho_N$, open circles) 
density distributions calculated in $^{90}{\rm Zr}+40\Lambda+\kappa K^-$ 
for $\kappa=0$ and $10$ $K^-$ mesons. The removal of the $1s$-state $\Xi$ 
hyperons from the primary baryonic configuration $^{90}{\rm Zr}+40\Lambda+ 
2\Xi^0+2\Xi^-$ affects considerably the hyperon density distribution $\rho_Y$ 
in the central region of the nucleus, this effect being magnified by the 
presence of $K^-$ mesons. In contrast, the nucleon density $\rho_N$ remains 
almost intact. For $\kappa=10$, $\Xi$ hyperons appear to repel nucleons from 
the center of the multi-$\{ N, Y, {\bar K}\}$ system, much like $\Lambda$ 
hyperons do.

\subsection{Multi-$\{ N, \Lambda, \Xi, K^+ \}$ configurations} 

\begin{figure} 
\includegraphics[scale=0.7]{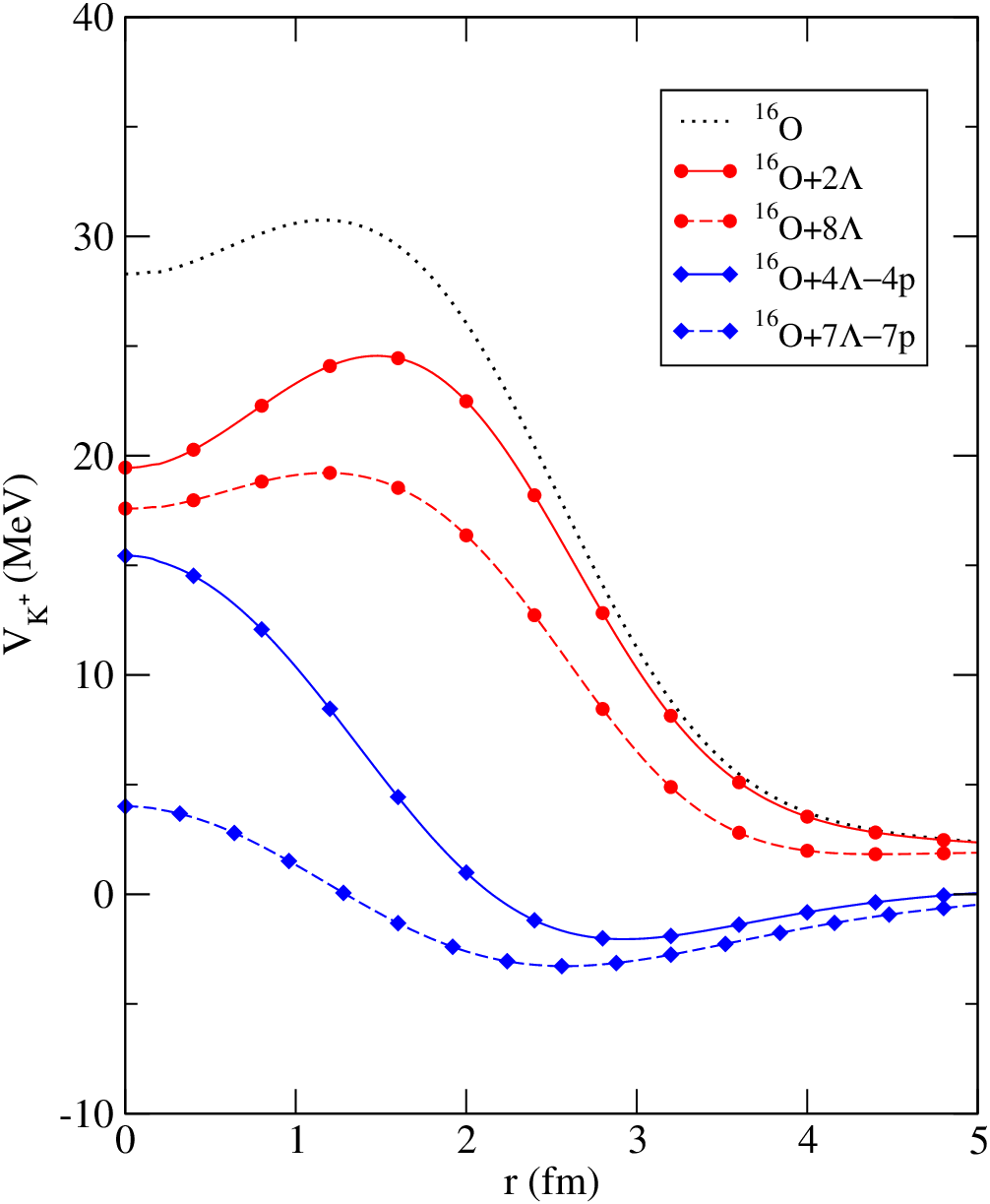} 
\caption{(Color online)The $K^+$ static potential in $^{16}{\rm O} + \eta 
\Lambda - \nu p$, calculated in the NL-SH RMF model.} 
\label{fig:f6} 
\end{figure} 

The $K^+$-nucleus potential is known to be repulsive, with 
V$_{K^+} \approx 30$~MeV at central nuclear density \cite{FGa07}. Schaffner 
and Mishustin \cite{SMi96} suggested that the presence of hyperons could lead 
eventually to a decrease of the repulsion that $K^+$ mesons undergo in nuclear 
matter so that the $K^+$ potential might even become attractive. Here we 
studied the possibility of binding $K^+$ mesons in hypernuclear matter, 
neglecting for simplicity dynamical effects arising from coupling $K^+$ 
mesons to the hypernuclear system. The $K^+$-nucleus potential was 
constructed simply by applying a $G$-parity transformation to the 
corresponding $K^-$ potential, choosing $g_{\sigma K}$ such that it produces 
$B_{K^-}=100$~MeV in the given core nucleus. 

Figure~\ref{fig:f6} shows the radial dependence of the real part of the static 
$K^+$ potential in various hypernuclear systems connected with $^{16}$O. The 
dotted line shows the repulsive $K^+$ potential in $^{16}$O for comparison. 
The figure indeed shows that the repulsion decreases, from roughly 30 MeV 
down to roughly 20 MeV with the number of $\Lambda$ hyperons added to the 
nuclear core, but the $K^+$ potential remains always repulsive in 
$^{16}$O+$\eta\Lambda$ systems. Searching for a $K^+$ bound state in hadronic 
systems we also calculated the $K^+$ potential in more exotic multistrange 
hypernuclei $^{\rm A}{\rm Z} + \eta\Lambda - \nu p$, where several protons are 
removed from the nuclear core in an attempt to increase the $|S|/B$ ratio and 
to reduce Coulomb repulsion. Figure~\ref{fig:f6} indicates that such removal 
of protons from $^{16}$O has a sizable effect on the shape of the $K^+$ 
potential, which may result in a shallow attractive pocket. However, the 
attraction is insufficient to bind a $K^+$ meson in these hadronic systems. 
Our calculations confirmed that the above conclusion holds also in heavier 
hypernuclear configurations based on Ca, Zr, and Pb cores. 

\begin{figure}
\includegraphics[scale=0.7]{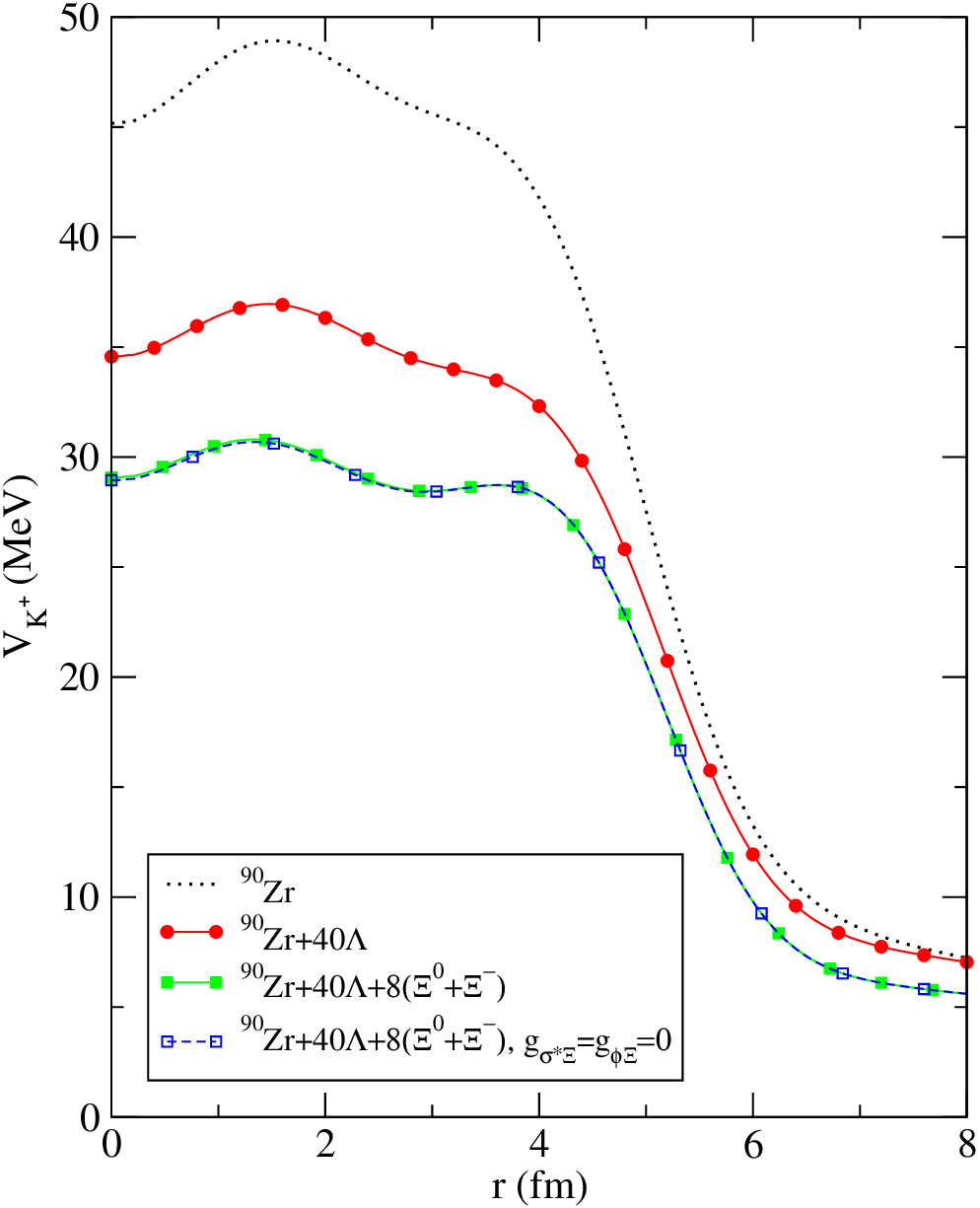} 
\caption{(Color online) The $K^+$ static potential in $^{90}{\rm Zr} + \eta 
\Lambda + \mu (\Xi^0 + \Xi^-)$, calculated in the NL-TM1 RMF model.} 
\label{fig:f7} 
\end{figure} 

In heavier nuclei, where it becomes possible to accommodate also $\Xi$ 
hyperons in addition to $\Lambda$ hyperons, the $K^+$ repulsion may be 
further reduced. This is demonstrated in Fig.~\ref{fig:f7} for a $^{90}$Zr 
nuclear core. However, this reduction is insufficient to reverse the repulsion 
into attraction. The figure also shows that the {\it hidden strangeness} 
couplings (chosen to be $g_{i\Xi}=2g_{i\Lambda},\; i=\sigma^*,\; \phi $) 
have no effect whatsoever on the reduction accomplished by the 
presence of $\Xi$ hyperons. 

Finally, we searched for $K^+$ bound states in nuclei sustained by $K^-$ 
mesons. The presence of deeply bound $K^-$ mesons makes the $K^+$ potential 
immensely deep (more than 100~MeV in $^{16}{\rm O}+8{\rm K}^-$). Hovever, 
because the $K^-$ mesons are concentrated at the very center of the nucleus, 
the $K^+$ potential is of a rather short range of about 1~fm. As a result, we 
found only very weakly bound $K^+$ states (by 1~MeV) in multi-$\{ N,Y,K^- \}$ 
configurations. A more careful treatment of $K^+K^-$ dynamics near threshold 
is necessary before coming to further conclusions, but our conclusion is not 
at odds with recent studies of the $I=1/2, J^{\pi}={1/2}^+$ $K{\bar K}N$ 
system \cite{JKE08,TKO09}.

\section{Summary and conclusions} 
\label{sec:fin} 

In this work, the RMF equations of motion for multi-$\bar K$ hypernuclei 
were formulated and solved for self-bound finite multistrange configurations. 
The choice of coupling constants of the constituents -- nucleons, hyperons, 
and $\bar K$ mesons -- to the vector and scalar meson fields was guided by 
a combination of accepted models and by phenomenology. The sensitivity to 
particular chosen values was studied. The results of the RMF calculations 
show a robust pattern of binding-energy saturation for $\bar K$ mesons as 
a function of their number $\kappa$. Compared to our previous RMF results 
for multi-$\bar K$ nuclei \cite{GFGM08}, the added hyperons do not bring 
about any quantitative change in the $B_{K^-}(\kappa)$ saturating curve. 
The main reason for saturation remains the repulsion induced by the vector 
meson fields, primarily $\omega$, between $\bar K$ mesons. 
The SU(3)$_V$ values adopted here for $g_{vK}$, Eq.~(\ref{eq:KSU(3)}), 
provide the ``minimal'' strength for $g_{\omega K}$ out of several other 
choices made in the literature, implying that the saturation of 
$B_{K^-}(\kappa)$ persists also for other choices of coupling-constant sets, 
as discussed in Ref.~\cite{GFGM08}. The repulsion between $\bar K$ mesons 
was also the primary reason for saturation in multi-$\bar K$ nuclei, both 
in our previous work \cite{GFGM08} and in Ref.~\cite{MMT09}. 

The saturation of $B_{K^-}$ with typical values below 200 MeV, considerably 
short of what it takes to replace a $\Lambda$ hyperon by a nucleon and a 
$\bar K$ meson, means that $\bar K$ mesons do not compete favorably and thus 
cannot replace hyperons as constituents of strange hadronic matter. In other 
words, $\bar K$ mesons do not condense in self-bound hadronic matter. 
The baryon densities of multi-$\bar K$ hypernuclei are between $(2-3)\rho_0$, 
where $\rho_0$ is nuclear-matter density. This is somewhat above the values 
obtained without $\bar K$ mesons, but still within the density range where 
hadronic models are likely to be applicable. 

Our conclusion of no ``kaon condensation'' is specific to self-bound finite 
hadronic systems run under strong-interaction constraints. It is not directly 
related to the Kaplan-Nelson conjecture of macroscopic kaon condensation 
\cite{KNe86}, nor to hadronic systems evolving subject to weak-interaction 
constraints, such as neutron stars. Yet, this conclusion has been challenged 
recently by Muto \cite{Muto08b} who uses the liquid-drop approach to claim 
that multi-$\bar K$ hypernuclei (termed by him ``kaon-condensed hypernuclei'') 
may provide the ground-state configuration of finite strange hadronic systems 
at densities about $9\rho_0$. Of course this high value of density for 
kaon-condensed hypernuclei is beyond the range of applicability of hadronic 
models, because quark-gluon degrees of freedom must enter in this density range. 
His calculation also reveals an isomeric multistrange hypernuclear state, 
without $\bar K$ mesons, at density about $2\rho_0$ which is close to what 
we find here within a RMF bound-state calculation. The appearance of 
a high-density kaon-condensed hypernuclear bound state in Muto's calculation 
might be just an artifact of the applied liquid-drop methodology, which does 
not provide an accurate substitute for a more microscopically oriented 
bound-state calculation. 

The role of $K^-$ strong decays in hadronic matter was played down in the 
present calculation of multi-$K^-$ hypernuclei because our aim, primarily, 
was to discuss and compare (real) binding energies of strange hadronic 
matter with and without $K^-$ mesons. The width of deeply bound $K^-$ 
nuclear configurations was explored by us in Refs.~\cite{MFG05,MFG06,GFGM07}, 
concluding that residual widths of order $\Gamma_{K^-} \sim 50$~MeV due to 
$K^-NN \to \Lambda N,\Sigma N$ pionless conversion reactions are expected 
in the relevant range of binding energy $B_{K^-} \sim 100 - 200$~MeV. 
This estimate should hold also in multi-$K^-$ hypernuclei where added 
conversion channels are allowed: $K^-NY \to \Lambda Y, \Sigma Y$, 
$K^-N \Lambda \to N \Xi$, and $K^- \Lambda Y \to \Xi Y$. We know of 
no physical mechanism capable of reducing substantially these widths, and 
therefore we do not anticipate multi-$K^-$ nuclei or multi-$K^-$ hypernuclei 
to exist as relatively long-lived isomeric states of strange hadronic 
matter which consists of multi-$\{ N,\Lambda,\Xi \}$ configurations.

\begin{acknowledgments} 
This work was supported in part by GACR grant 202/09/1441 and by SPHERE within
the FP7 research grant system. AG acknowledges instructive discussions with
Wolfram Weise and the support extended by the DFG Cluster of Excellence, Origin
and Structure of the Universe, during a visit to the Technische Universit\"{a}t
M\"{u}nchen. 
\end{acknowledgments}

\end{document}